\documentclass{pasj00}

\newcommand{\arcm}{\hbox{$^\prime$}}
\newcommand{\arcs}{\arcm\hskip -0.1em\arcm}

\begin{document}
\SetRunningHead{Andrew Read \etal}{First Results from the XMM-Newton Slew Survey}
\Received{2000/12/31}
\Accepted{2001/01/01}

\title{First Results from the XMM-Newton Slew Survey}


 \author{%
   Andrew \textsc{Read}\altaffilmark{1}
   Richard \textsc{Saxton}\altaffilmark{2}
   Pilar \textsc{Esquej}\altaffilmark{2,3}
   Michael \textsc{Freyberg}\altaffilmark{3}
   and
   Bruno \textsc{Altieri}\altaffilmark{2} }
 \altaffiltext{1}{Dept. of Physics and Astronomy, Leicester University, Leicester LE1\,7RH, U.K.}
 \email{amr30@star.le.ac.uk}
 \altaffiltext{2}{European Space Agency (ESA), European Space Astronomy Centre, Villafranca, 28080 Madrid, Spain}
 \altaffiltext{3}{Max-Planck-Institut f\"ur extraterrestrische Physik, 85748 Garching, Germany}


\maketitle

\begin{abstract}

We have attempted to analyse all the available data taken by
XMM-Newton as it slews between targets. This slew survey, the
resultant source catalogue and the analysis procedures used are
described in an accompanying paper. In this letter we present the
initial science results from the survey. To date, detailed
source-searching has been performed in three X-ray bands (soft, hard
and total) in the EPIC-pn 0.2$-$12\,keV band over $\sim$6300
sq.degrees ($\sim$15\% of the sky), and of order 4000 X-ray sources
have been detected ($\sim$55\% of which have IDs). A great variety of
sources are seen, including AGN, galaxies, clusters and groups, active
stars, SNRs, low- and high-mass XRBs and white dwarfs. In particular,
as this survey constitutes the deepest ever hard-band 2$-$12\,keV
all-sky survey, a large number of hard sources are
detected. Furthermore, the great sensitivity and low-background of the
EPIC-pn camera are especially suited to emission from extended
sources, and interesting spatial structure is observed in many
supernova remnants and clusters of galaxies. The instrument is very
adept at mapping large areas of the X-ray sky. Also, as the slew
survey is well matched to the ROSAT all-sky survey, long-term
variability studies are possible, and a number of extremely variable
X-ray sources, some possibly due to the tidal disruption of stars by
central supermassive black holes, have been discovered.

\end{abstract}

\section{Introduction}

XMM-Newton (Jansen \etal\ 2001), with the huge collecting area of its
mirrors and the high quantum efficiency of its EPIC detectors
(Str\"{u}der \etal\ 2001, Turner \etal\ 2001), is the most sensitive
X-ray observatory ever flown. This is strikingly evident during slew
exposures which, while yielding only at most around 15 seconds of
on-source observing time, actually constitute a hard-band 2-10 keV
survey with a limiting flux sensitivity of
$\approx4\times10^{-12}$\,erg cm$^{-2}$ s$^{-1}$. This is 5$-$10 times
deeper than all other all-sky surveys. The soft-band 0.2-2 keV survey
(limiting flux sensitivity $\approx6\times10^{-13}$\,erg cm$^{-2}$
s$^{-1}$) is comparable with the ROSAT PSPC all-sky survey (RASS),
offering long-term variability studies.

\section{First Results from the XMM-Newton Slew Survey}

With the release of the first source catalogue (XMMSL1, released in
May 2006) into the XMM X-ray Science Archive -
\verb+http://xmm.vilspa.esa.es/+
\verb+external/xmm_data_acc/xsa/index.shtml+, we are able here to show
selected highlights seen so far. This will allow the
community to see immediately what science is possible using the
Slew Survey data and catalogue.
General results (catalogue properties, source population properties
etc.) from the XMM-Newton Slew Survey first catalogue (XMMSL1) are
described in full, along with a description of how the analysis of the
slew data was performed and how the catalogue was constructed, in
Saxton \etal\ (in preparation). In the following subsections we
present a concise report of the first major science results from the
XMM-Newton Slew Survey:

\subsection{ Identifications and high redshift objects}

All of the catalogued detected slew sources were cross-correlated with
several catalogues and databases, including Simbad, NED, HEASARC, RASS
etc. Counterparts were found for over 50\% of the sources, and these
counterparts can be found in the XMMSL1 slew catalogue. 17\% of the
non-extended slew sources are correlated (within 30\arcsec) with
sources in the Sloan Digital Sky Survey (SDSS) DR5 imaging
catalogue. This is 86\% of the possible counterparts, given the
isotropic distribution of the slew sources and the area of the SDSS
DR5 area (8000 sq.degrees) $-$ i.e. most of the slew sources within
the SDSS area have SDSS matches. Very large numbers (several hundred)
of AGN and galaxies were found in the Slew Survey, the most distant
being a QSO (J0646+4451) (or HB89~0642+449) at a redshift of
z=3.4. Its count rate of 1.2\,ct~s$^{-1}$ corresponds at this redshift
to a luminosity of 8.5$\times10^{46}$\,erg s$^{-1}$ (assuming a
$\Gamma=1.7$ power law and $H_{0}$=75\,km s$^{-1}$ Mpc$^{-1}$).

\subsection{ Hardness ratios}
 
Over 150 of the detected sources are detected both in the soft and the
hard band, and we are therefore able to construct hardness ratios for
these sources. Plotting the spectral hardness against the Galactic
latitude, we see that the very hardest sources all lie at a Galactic
latitude of $\sim$0, i.e. they lie in the Galactic plane. Of these
sources, the ones that we have been able to identify are all LMXRBs.
Conversely, hardly any soft sources at all are seen in the Galactic
plane, and of the few that are, none are extragalactic. This is as
expected due to Galactic absorption, but is very encouraging to
observe.

\subsection{ Extremely bright sources}
 
\begin{figure}
\centering
\includegraphics[bb=100 70 550 900,clip,width=6.0cm,angle=270]{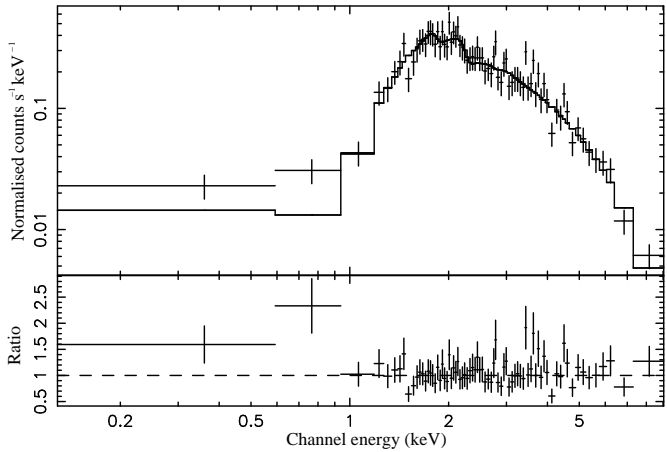}
\caption{XMM-Newton Slew Survey spectrum (from $\approx8$\,sec of
exposure time) of V2216~Oph (4U 1728-16), a believed LMXRB source
towards the Galactic Centre. An absorbed high$-kT$ bremsstrahlung
model has been fitted to the data, and the data$-$model residuals are
shown in the lower panel.}
\label{fig_spec}
\end{figure}

A number of extremely bright objects have been detected in the slew
survey, including several LMXRBs close to the Galactic Centre. Three
of the brightest detected are 4U~1758-25, SGR~X-3 \& V2216~Oph, all
known ROSAT sources, with EPIC-pn count-rates of between 330 and
530\,ct s$^{-1}$. Another extremely bright source detected in the slew
is a potential black hole candidate, previously seen not by ROSAT, but
by XTE $-$ XTE\_J1746-319. The brightest non-Galactic sources detected
are in the LMC $-$ LMC X-2 and LMC X-3. Note that these extremely
bright sources are severely affected by pile-up, and their true
count-rates would be much higher in optimum (smaller frame-time)
observing conditions. This fact, along with the as yet unsolved
problems associated with sources moving swiftly across the detector,
and the associated response matrix and effective area/PSF issues make
quantitative spectral analysis extremely difficult at
present. However, spectra can be extracted from the brightest sources,
and rudimentary spectral analyses give satisfactory results. The
spectrum extracted from V2216~Oph (4U 1728-16) for example
(Fig.\ref{fig_spec}) appears wholly consistent with the source being
an LMXRB, as is believed.

\subsection{ Hard-band (2--10\,keV) sources}
 
As the hard-band 2$-$10\,keV slew survey is 5$-$10 times deeper than
all other previous and current all-sky surveys, it is no surprise that
a large number of sources are detected in this band. Comparing our
results with those of the HEAO-1 A2 All Sky Survey (Piccinotti \etal\
1982), we see that whereas they detect 85 sources in the high latitude
sky (excluding the SMC and the LMC) in an area totalling 65\% of the
sky, we detect 148 hard-band sources in similar sky regions (though
covering only 8\% of the sky). In the hard X-ray band, therefore, the
XMM-Newton slew survey sees approximately 14 times more sources per
unit area than the HEAO-1 A2 All Sky Survey, and this will give very
useful input to the 2$-$10 keV Log-N/Log-S curve in the few times
10$^{-12}$ to the few times 10$^{-11}$\,erg cm$^{-2}$ s$^{-1}$ range.
Issues of Eddington bias and completeness will need a careful
treatment though (see Saxton \etal\ (in preparation)).

\subsection{ Multiple detections}
 
Some of the XMM-Newton slews analysed to date are seen to recross
areas of the sky covered by previous slews. As such the possibility
exists to go to fainter fluxes in certain areas of the sky, as more
and more slews pass over the same area. Also, variability studies
within the slew survey are possible. 105 of the sources within the
XMMSL1 catalogue have been seen in more than one slew, including the
black hole candidate XTE\_J1746-319 $-$ in this particular case, no
evidence for variability in flux or spectrum is seen. Conversely, a
source detected in three separate slews associated with the galaxy
2MASXJ06100652-6243125, shows evidence for a factor $\sim$2 variation
in flux in only 10\,days.

\subsection{ Correlations with ROSAT}
 
Cross-correlating the XMM-Newton slew catalogue with the ROSAT
all-sky-survey catalogue (RASS; Voges \etal\ 1999)  reveals that
approximately 50\% of the non-extended slew sources (with detection
likelihood $>10$) have RASS counterparts within 30\arcs. Many of the
remaining sources are either X-ray hard (and perhaps invisible to
RASS), are variable, or are in the small portions of the sky not
covered by the RASS.

\subsection{ Variable sources}
 
\begin{figure}
\centering
\includegraphics[bb=40 150 577 663,clip,width=8.5cm]{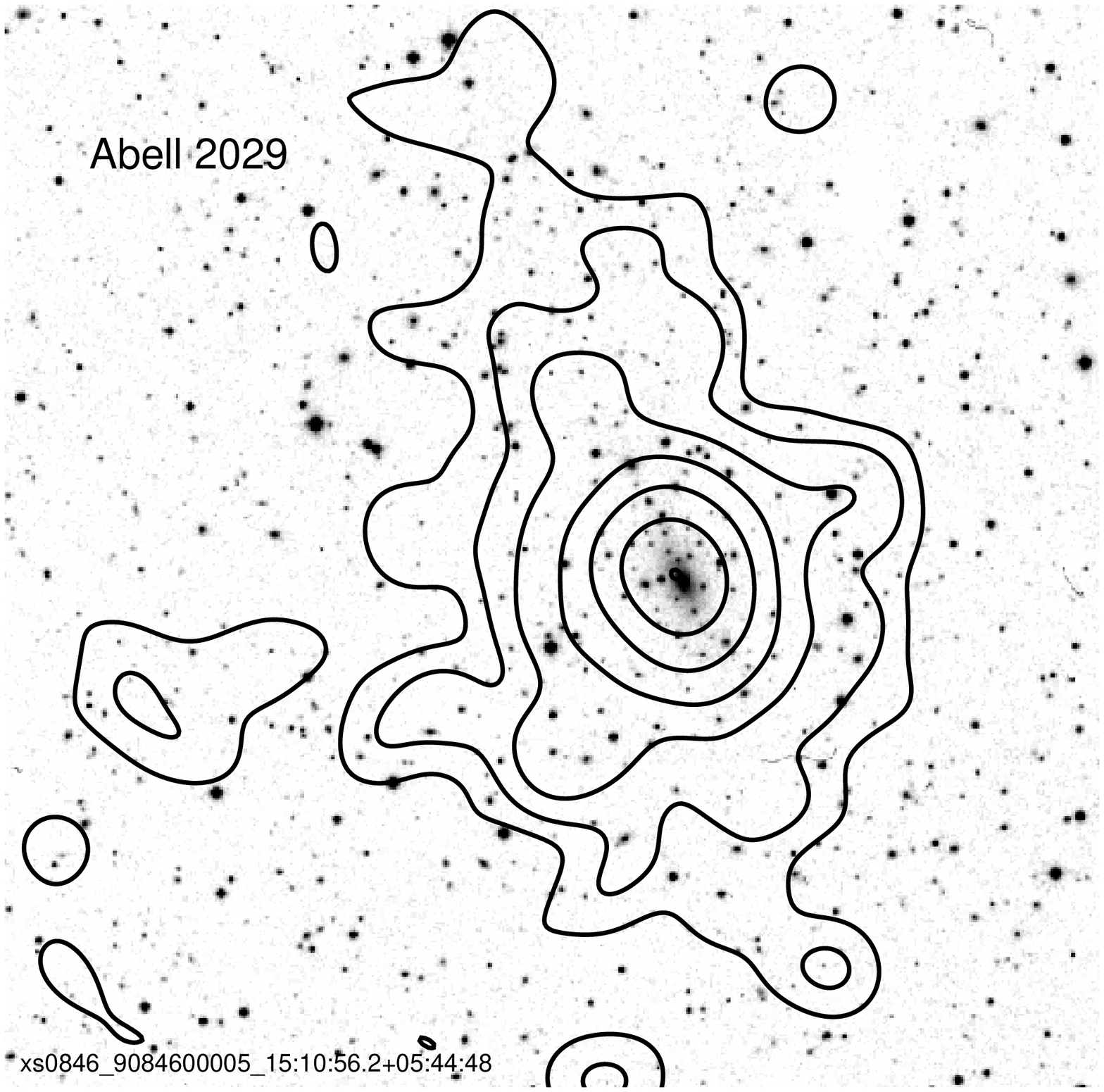}
\includegraphics[bb=70 318 546 427,clip,width=8.5cm]{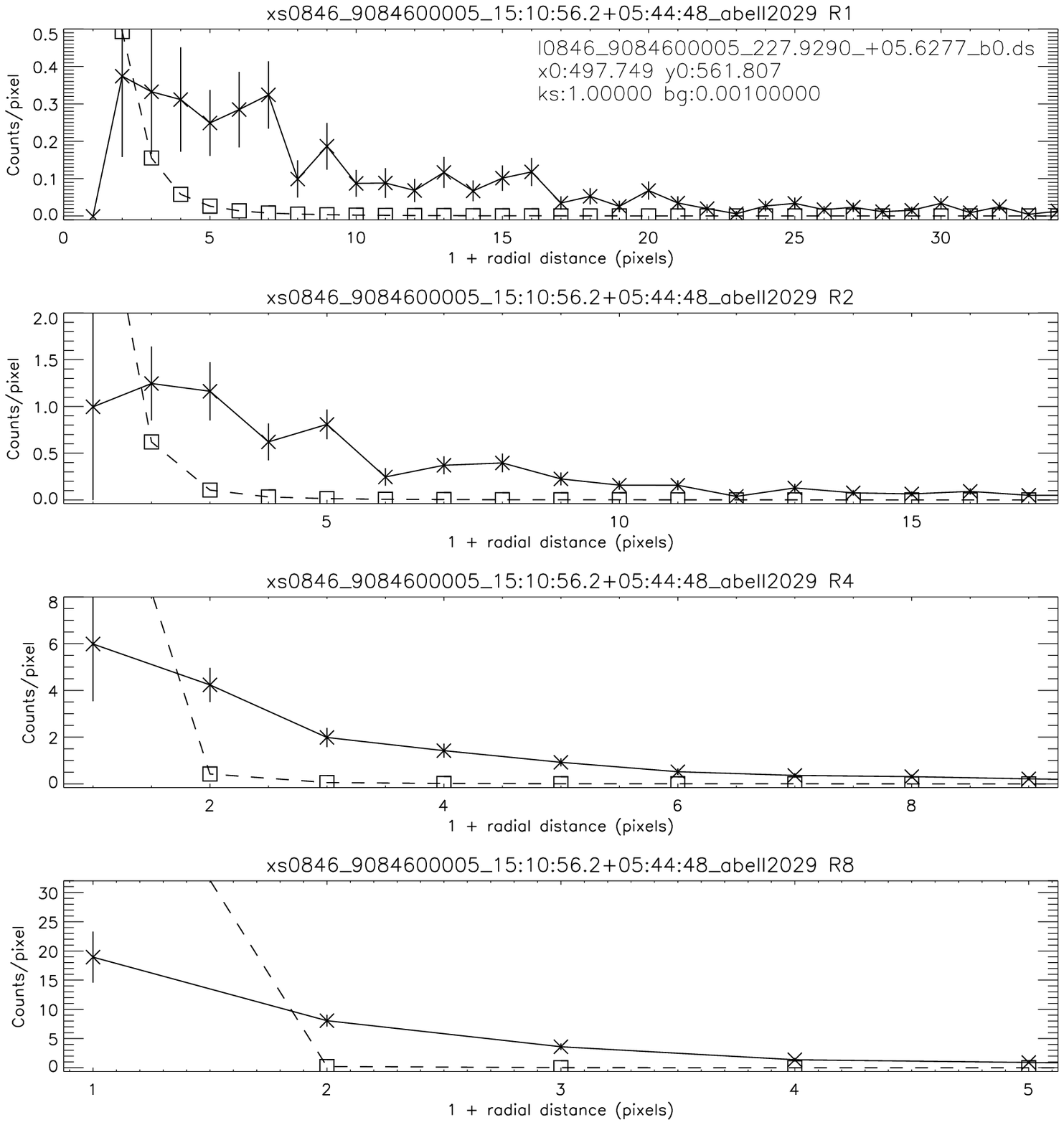}
\caption{(Top) Contours of (0.2$-$12\,keV) XMM-Newton Slew X-ray
emission (exposure approximately 10\,sec) superimposed on a DSS image of
the galaxy cluster Abell 2029. (Bottom) A radial profile of the
XMM-Newton Slew X-ray emission from Abell 2029, compared with that
from a bright point source.}
\label{fig_cluster}
\end{figure}

Approximately 1\% of the sources with RASS counterparts show a
RASS-to-slew flux variability greater than a factor of 10. Many of
these are perhaps expected. The variable star UY~Vir, for instance,
shows a RASS-to-slew flux increase by a factor of $\approx$25, whereas
the Cataclysmic Variable QQ~Vul shows a decrease by about the same
factor.

Of great interest though, are the sources showing extreme variability
where no variability was expected. A slew source coincident with the
otherwise innocuous, small, elliptical, non-active galaxy NGC3599 is
seen to have increased in flux by about a factor of almost 200 since
the RASS. The fact that this source is very soft allows us to consider
such a huge variability in terms of flaring radiation emitted due to
the tidal disruption of a star by a supermassive quiescent black hole
at the centre of the galaxy. This source and other similar
high-variability, possible tidal disruption events, are discussed
fully in Esquej \etal\ (in preparation). Note that it is only when a
large fraction of the sky is observed, as is the case here, that rare
events such as these have a chance of being observed.

\subsection{ Extended sources}
 
Approximately 15\% of the catalogued sources show some significant
extension beyond the PSF, and a number of these have been identified
with all manner of extended sources including supernova remnants,
galaxies and galaxy groups and clusters: 

\subsection{ Galaxies}
 
The cross-correlations reveal that the vast majority of objects that
are seen (for which we have an object type) are galaxies, and AGN make
up approximately half of the 900 or so cross-correlations with known
galaxies. Several NGC and Markarian galaxies are seen, including a
number of starburst galaxies (notably M83) and interacting galaxies
(an example being NGC4748). A large number of Seyferts (1 and 2),
BLLACs and QSOs (extending out to a redshift of $z=3.4$) are also
detected.

\subsection{ Galaxy Groups and Clusters}
 
Of the catalogued sources that show some significant
extension, many are associated with known groups and clusters. A total
of 81 Abell and Zwicky clusters have apparent slew counterparts, and a
great many of these are seen to have large extension values. Indeed
these Abell/Zwicky sources make up a large fraction (approximately
half) of all the slew sources which have large values of estimated
extent. In a number of the brighter cases, the size and extent of the
cluster can be mapped, both on the sky and as a radial profile (see
Fig.\,\ref{fig_cluster}). Sources associated with galaxy groups have
also been detected, two examples being the NGC1566 and NGC2300 groups.

The XMM-Newton Slew Survey is seen to be very sensitive to relatively
faint clusters and extended sources in general thanks to the wide
band-pass, the tight PSF and the low background. At present, the
estimated average cluster sensitivity limit for the slew survey is
just over $10^{-12}$\,erg s$^{-1}$ cm$^{-2}$ (at 0.2$-$2\,keV),
calculated for a $N_{H}=3\times10^{20}$\,cm$^{-2}$, $kT=5$\,keV
cluster. This is comparable with that of the MACS survey (Ebeling
\etal\ 2001), though the sky area covered ($\sim$15\% of the sky) is
much smaller than that of MACS. As the Slew Survey progresses however,
the area of sky covered will increase to perhaps that of MACS and
beyond, where more than 100 high redshift ($z>0.3$) X-ray luminous
clusters should be observable.

\subsection{ Supernova Remnants}
 
A great many sources that were initially formally detected as
individual sources were seen, once careful cross-correlations were
performed, to be instead due to large Galactic extended sources,
i.e.\,supernova remnants (SNR). For example, a great number of these
spurious point sources were seen associated with Puppis~A. This was
not merely confined to our own Galaxy, as many other sources were seen
associated with other SNR in the LMC (such as N132D), and in the SMC.

   \begin{figure*}
   \centering
   \includegraphics[bb=35 255 575 545,clip,width=18cm]{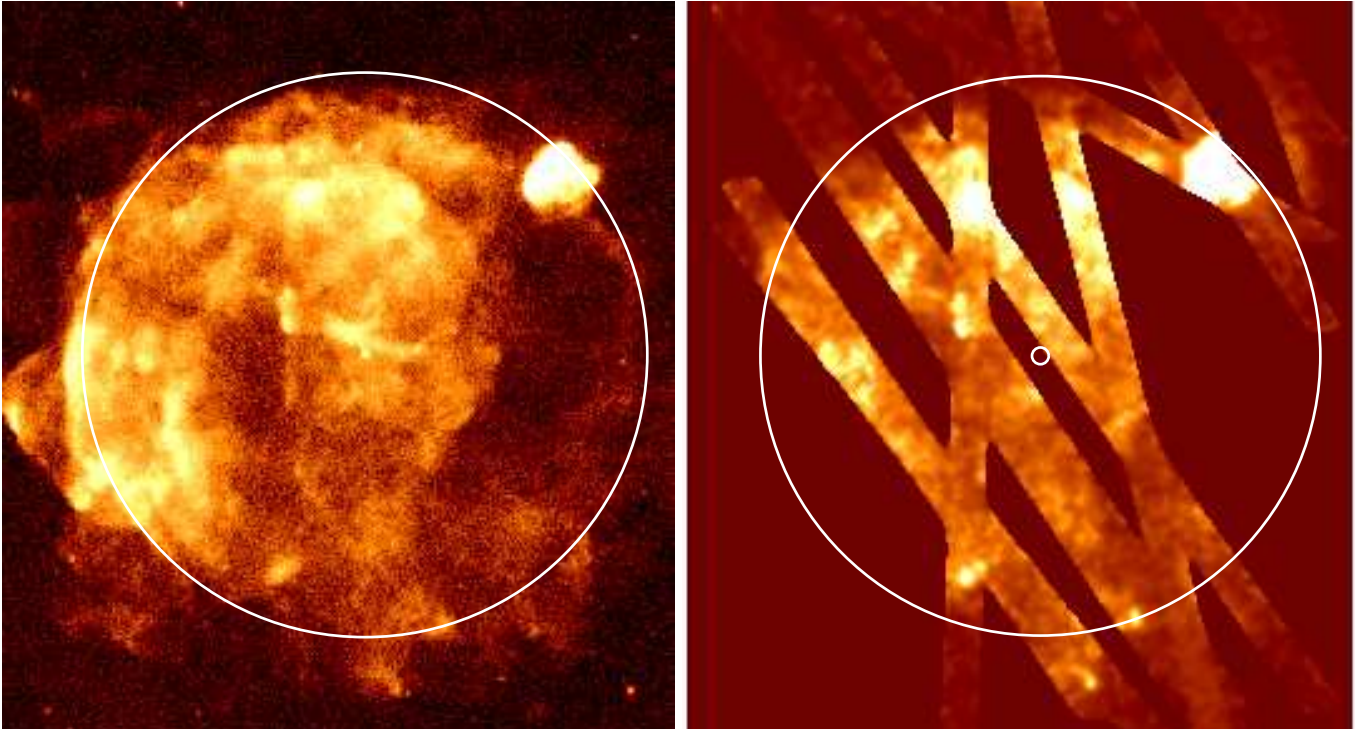}
\caption{(Left) ROSAT PSPC all-sky survey (0.1$-$2.4\,keV)
exposure-corrected image of the Vela SNR (Aschenbach \etal\
1995). (Right) Current XMM-Newton Slew lightly-smoothed,
exposure-corrected (0.2$-$12\,keV) image of the same area of sky. A
3.5$^{\circ}$ radius circle is marked on both images to guide the eye
(also a small circle is marked on the XMM-Newton Slew image indicating
the position of the Vela PSR). }
\label{fig_vela}
    \end{figure*}

\subsection{ Very extended sources and large-scale features}
 
As a large percentage of the whole sky has currently been covered by
the XMM-Newton slew (the released catalogue covers $\sim$6300
sq.degrees), and certain areas have been covered more than once, then
we would expect certain sky areas to have quite a large
coverage. Given the orbit of XMM-Newton, the slewing strategy is such
that a great many of the slews pass close to (though not directly
over) the north and south ecliptic poles. One is able to combine many
slew images together to obtain large-scale maps, and an example of
this is shown in Fig.\ref{fig_vela}, where a ROSAT PSPC image of the
Vela SNR is shown compared with the current XMM-Newton Slew
large-scale image of the same field. Vela, together with Puppis~A, the
bright, extended ($\approx$1$^{\circ}$ diameter) SNR at the
north-western edge of Vela, lie fairly close to the region where a
large number of slews are seen to cross. The image shows that only a
few seconds of XMM-Newton Slew time are necessary to observe the
features that are seen in the ROSAT image. As the mission progresses
more slews will become available to add to this picture. In addition
to this, XMM-Newton slew data exists not only in the (ROSAT-like) soft
X-ray band (0.2$-$2\,keV), but also in the hard X-ray (2$-$12\,keV)
band. Using this hardness information, we see that Puppis~A appears
somewhat harder than the vast majority of Vela, though a small number
of harder features are also seen within the main Vela remnant. The
potential therefore for XMM-Newton being able to map out large areas
of the X-ray sky via slews such as these is excellent.

\section{Conclusions}

XMM-Newton slew exposures yield at most only 15 seconds of on-source
observing time. The high quantum efficiency of its EPIC-pn detector
plus the huge collecting area of its mirrors however, result in an
XMM-Newton Slew hard-band (2$-$10\,keV) survey ten times deeper than
all other all-sky surveys, and a soft-band (0.2$-$2\,keV) survey
comparable with the ROSAT PSPC all-sky survey.

With the release of the XMM-Newton Slew Survey first source catalogue
(XMMSL1) and data files into the XMM X-ray Science Archive, we have
been able here to show the first science results from the
survey. These include:

  \begin{itemize}

\item Identifications of many sources, including a large number of
galaxies and AGN, with the current most distant XMM-slew object (a
QSO) lying at a redshift of $z=3.4$.

\item Several extremely bright sources, some not seen by ROSAT, both
in our Galaxy, especially close to the Galactic Centre, and beyond.

\item A large number of hard (2$-$10\,keV) sources $-$ The XMM-Newton
slew survey is the deepest ever hard-band all-sky survey, and sees
$\approx$14 times more sources per unit area than the HEAO-1 A2
All-Sky Survey. 

\item Sources detected in more than one slew, some showing
variability.

\item Extremely variable sources, when compared with the RASS,
some possibly due to the tidal disruption of stars by central supermassive
black holes (Esquej et al., in preparation).

\item Several detections of extended sources, including galaxies,
galaxy groups and especially galaxy clusters, plus also more nearby
supernova remnants in our Galaxy and in the Magellanic clouds.

\item The emerging excellent ability of XMM-Newton to map large areas
of the X-ray sky, by co-adding several slews together. The example of
the Vela SNR (plus Puppis~A) indicates that a few seconds of
XMM-Newton time are all that are needed to reveal the X-ray features
observed with ROSAT.

   \end{itemize}

\end{document}